\documentclass{iau} 
\usepackage{graphicx}
\usepackage{natbib}

\title[Dynamical masses of Cepheids from the GAIA parallaxes] 
{Dynamical masses of Cepheids from the GAIA parallaxes}

\author[Gallenne et al.]   
{A.~Gallenne$^1$, P.~Kervella$^2$, A.~M\'erand$^3$, N.~R.~Evans $^4$
 \and  C.~Proffitt$^5$}

\affiliation{$^1$European Southern Observatory, Alonso de C\'ordova 3107, Casilla 19001, Santiago 19, Chile \\[\affilskip]
$^2$LESIA, Obs. de Paris, CNRS UMR 8109, UPMC,Univ. Paris 7, 5 Pl. Jules Janssen, 92195 Meudon, France \\[\affilskip]
$^3$European Southern Observatory, Karl-Schwarzschild-Stra\ss e 2, 85748 Garching bei M\"unchen, Germany\\[\affilskip]
$^4$Smithsonian Astrophysical Observatory, MS 4, 60 Garden Street, Cambridge, MA 02138, USA \\[\affilskip]
$^5$Space Telescope Science Institute, 3700 San Martin Drive, Baltimore, MD 21218, USA \\[\affilskip]
}

\pubyear{2017}
\volume{xxx}  
\jname{Astrometry and Astrophysics in the Gaia sky}
\editors{A.~Recio-Blanco, P.~de Laverny, A.~Brown \& T.~Prusti eds.}
\begin{document}

\maketitle

\begin{abstract}
The mass of a Cepheid is a fundamental parameter for studying the pulsation and evolution of intermediate-mass stars. But determining this variable has been a long-standing problem for decades. Detecting the companions (by spectroscopy or imaging) is a difficult task because of the brightness of the Cepheids and the close orbit of the components. So most of the Cepheid masses are derived using stellar evolution or pulsation modeling, but they differ by 10-20\,\%. Measurements of dynamical masses offer the unique opportunity to make progress in resolving this mass discrepancy.

The first problem in studying binary Cepheids is the high contrast between the components for wavelengths longer than $0.5\,\mu$m, which make them single-line spectroscopic binaries. In addition, the close orbit of the companions ($<40$\,mas) prevents us from spatially resolving the systems with a single-dish 8m-class telescope. A technique able to reach high spatial resolution and high-dynamic range is long-baseline interferometry. We have started a long-term program that aims at detecting, monitoring and characterizing physical parameters of the Cepheid companions. The GAIA parallaxes will enable us to combine interferometry with single-line velocities to provide unique dynamical mass measurements of Cepheids.

\keywords{techniques: interferometric, stars: Cepheids, binaries}
\end{abstract}

\firstsection 
\section{Introduction}

Cepheids are powerful astrophysical laboratories providing fundamental clues for studying the pulsation and evolution of intermediate-mass stars. However, one of the most critical parameters, the mass, is a long-standing problem because of the 10-20\,\% discrepancy between masses predicted from stellar evolution and pulsation models. Cepheids in binary systems are the only tool to constrain models and make progress on this discrepancy. Studying Cepheid companions can also provide insight on the impact of binarity on the calibration of the period-luminosity relation from the Baade-Wesselink technique and the IR surface-brightness method. Detectable companions are mostly hot main-sequence stars, therefore the flux contribution in the near-IR might is often negligible ($<$\,1-2\%), but can be as large as 10\,\% in $V$.

Most of known companions are too close to the Cepheid ($<$\,40 mas) to be resolved with single-dish 8m class telescopes. In addition, the high-contrast between the components makes the detection even more difficult. But long-baseline interferometry (LBI) is able to reach high-spatial resolution and high-dynamic range, which allow us to accurately determine the astrometric position of some companions.

\section{Combining spectroscopy, interferometry and GAIA}
\begin{figure*}
	\centering
	\resizebox{\hsize}{!}{\includegraphics{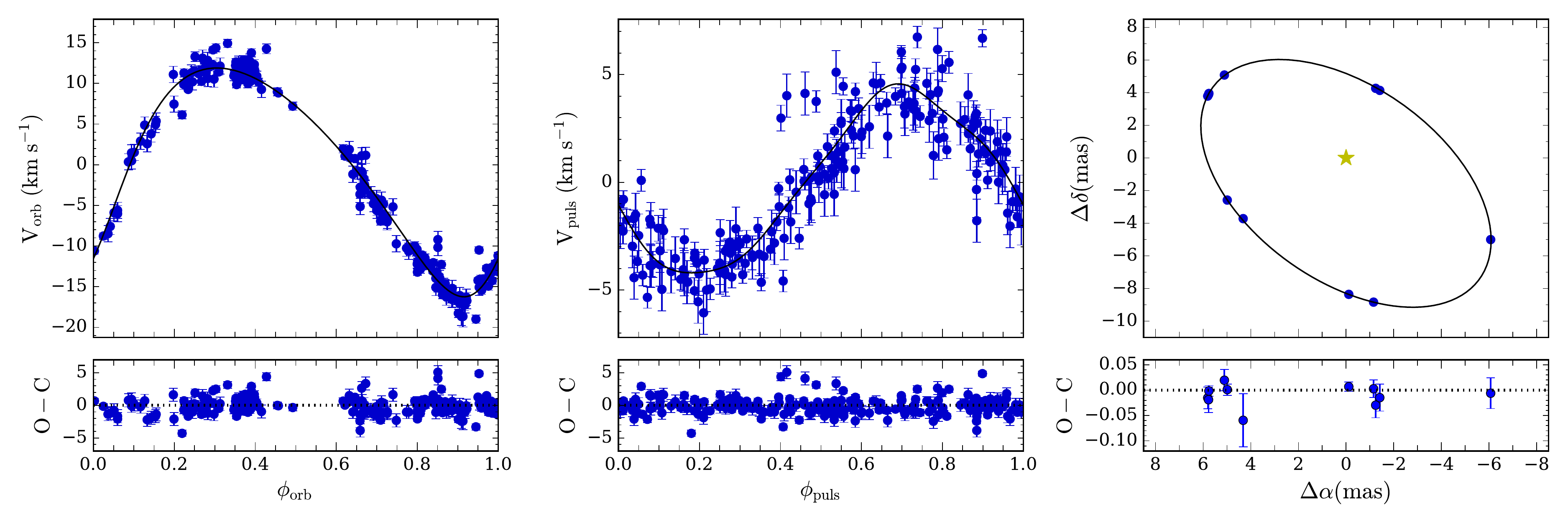}}
	\caption{Combined fit of single-line velocities with astrometry for the Cepheid V1334~Cyg.  Left: orbital velocity. Middle: pulsation velocity. Right: astrometric motion of the companion.}
	\label{fig:v1334}       
\end{figure*}

Because of the brightness of the Cepheid, it is challenging to spatially/spectrally detect companions using ground-based telescopes. With spectroscopy in the V band, the companion’s lines are difficult to disentangle from the Cepheid’s, mainly because of the high-contrast. However, the more massive companion causes the Cepheid to wobble (as the two components orbit their center of mass). This is more easily detectable from radial velocities (RVs) of the Cepheids, which makes them single-line spectroscopic binaries. From space, ultraviolet observations from HST/STIS can provide the spectra of the companions. However, broad features (some of them are fast rotators) and blended lines complicate the analysis, and can prevent the determination of accurate RVs. Combining astrometry and RVs of both components will yield estimates on the masses and distances.

With interferometry, we have observed some binary Cepheids in both the northern and southern hemisphere. We detected companions separated by less than 50\,mas from the Cepheids with contrast as high as 6.5\,mag in $H$. For this purpose, we have created a dedicated tool, named \texttt{CANDID}, which searches for high-contrast companions from interferometric data \citep{Gallenne_2015_07_0}. The tool delivers, among other things, the flux ratio $f$, the relative astrometric separation $(\Delta\alpha, \Delta\delta)$, and the (non-)detection level of the companion based on $\chi^2$ statistics. So far, we have detections for six binary Cepheids  (V1334~Cyg, AW~Per, RT~Aur, BP~Cir, AX~Cir, S~Mus), with projected separations ranging from 1.5 to 40\,mas and flux ratios (in $H$) from 0.8 to 4\,\%. The astrometric positions can be combined with the RVs of the primary, as shown in Fig.\ref{fig:v1334} for V1334~Cyg, to provide the full set of orbital elements, including $a, i$ and $\Omega$ previously unknown \citep{Gallenne_2013_04_0,Gallenne_2014_01_0}. However, the distance and masses are still degenerate, and unfortunately no accurate parallaxes exist for these binary Cepheids. The coming GAIA parallaxes will allow us to break this degeneracy, and we will then be able to combine interferometry with single-line velocities to provide dynamical mass measurements of Cepheids.

\end{document}